\documentclass[a4paper,11pt]{article}
\usepackage{pos}
\usepackage{slashed}

\title{Finite-volume effects and meson scattering in the 2-flavour Schwinger model}

\ShortTitle{Finite-volume effects and meson scattering in the 2-flavour Schwinger model}

\author*[a]{Patrick B\"uhlmann}
\author[a]{Urs Wenger}

\affiliation[a]{Albert Einstein Center for Fundamental Physics\\
    Institute for Theoretical Physics\\
    University of Bern\\
    Sidlerstrasse 5\\
    CH–3012 Bern\\
    Switzerland}

\emailAdd{ventura@itp.unibe.ch}
\emailAdd{wenger@itp.unibe.ch}

\abstract{We investigate the 2-flavour Schwinger model in the
  canonical formulation with fixed fermion numbers. We use Wilson
  fermions and a formalism which describes the determinant of the
  Dirac operator in terms of dimensionally reduced canonical transfer
  matrices. These transfer matrices allow the direct examination of
  arbitrary multi-particle (meson) sectors and the determination of
  the corresponding ground-state energies. We discuss the
  finite-volume effects in the meson mass.  From the $2$-meson
  energies, we determine the scattering phase shifts and compare the
  $3$-meson energies at finite volume to predictions based on
  3-particle quantization conditions.
}

\FullConference{%
 The 38th International Symposium on Lattice Field Theory, LATTICE2021
  26th-30th July, 2021
  Zoom/Gather@Massachusetts Institute of Technology
}

\begin{document}
\maketitle
\section{Introduction}
The Schwinger model \cite{Schwinger:1962tp} is of great interest since
it shares many similarities with Quantum Chromodynamics (QCD), such as
confinement, chiral symmetry breaking, charge shielding, and a
topological $\theta$-vacuum \cite{COLEMAN1975267,COLEMAN1976239}.
Thanks to these similarities the model is often used as a toy model to
test new computational strategies. In our case, we perform numerical
computations in the $2$-flavour Schwinger model using the canonical
formulation.  The corresponding canonical partition functions allow
one to consider the physics of the system in sectors with a fixed
number of particles, i.e., with fixed numbers of fermions, and to
determine the corresponding ground-state energies.  Using appropriate
ratios of the canonical partition functions we directly access the
energy spectrum of (multi-)meson states without resorting to the
computation of correlation functions. These become more and more
complicated with an increasing number of mesons
\cite{Detmold:2008fn}. In contrast, the complexity for the computation of the
partition functions is independent of the number of mesons involved.

We use the (multi-)meson ground-state energies to perform some
meson-scattering analysis. In the 2-flavour Schwinger model, the
canonical sectors with fixed fermion numbers are characterized by
their isospin content. Consequently, the corresponding states with the
lowest energies are the $n$-meson states of maximal isospin.  The
lowest-lying energies in the isospin $I=1$ sector, for example,
describe single-meson energies, while the lowest-lying energies in the
isospin $I=2$ sector correspond to the energies of the 2-meson
scattering states of maximal isospin, and so forth. For the
investigation of the meson scattering, we determine the mass of the
isospin $I=1$ meson (corresponding to the pion in QCD) on a large
range of spatial volumes in order to control the finite-volume
effects. Then, we calculate the scattering phase shifts from the
energies of the 2-meson states (corresponding to 2-pion scattering
states in QCD). Finally, from the isospin $I=3$ sector, we determine
the 3-meson energies and compare them to predictions from quantization
conditions for 3-particle energies based on the 2-meson scattering
phase shift \cite{Guo:2016fgl,Guo:2018xbv}.

The computation of the canonical partition functions is based on the
dimensional reduction of the fermion determinant in terms of transfer
matrices \cite{Alexandru:2010yb}. Using those, it is then
straightforward to project onto the canonical determinants describing
the dynamics of the fermions in the sectors with fixed fermion numbers
\cite{Steinhauer:2014oda}.  In the context of QCD, the canonical
formulation has been used with staggered and Wilson fermions, see
Refs.~\cite{Hasenfratz:1991ax,Kratochvila:2005mk,Fodor:2007ga} and
\cite{Alexandru:2005ix}, respectively, for some early applications. In
some cases, the canonical formulation is also useful to solve fermion
sign problems, see
Refs.~\cite{Alexandru:2017dcw,Burri:2019wge,Buhlmann:2021gzz}. Here we
consider the $2$-flavour Schwinger model with Wilson fermions at fixed
isospin density where the fermion sign problem is not present.

\section{The 2-flavour Schwinger model in the canonical formulation}
Using the doublet $\psi = (u, d)$ to describe the two flavours of
mass-degenerate fermions with opposite isospin charges (in
correspondence with the up and down quarks in QCD), the continuum
Lagrangian $\mathcal{L}$ of the $2$-flavour variant of the Schwinger
model is given by
\begin{align}\label{eq:LagrangianSchwingerModel}
	\mathcal{L}[\bar{\psi},\psi,A_{\mu}] = \bar{\psi}(x)[i \slashed{D}-m_{0}]\psi(x)-\frac{1}{4}F_{\mu \nu}F^{\mu \nu},
\end{align}
where $D_{\mu} = \partial_{\mu}+ig A_{\mu}(x)$ is the covariant
derivative with the Abelian gauge field $A_{\mu}(x)$, $m_0$ the mass
of the two fermions, $g$ the gauge coupling and
$F_{\mu \nu}(x) = A_{\mu}(x)A_{\nu}(x)-A_{\nu}(x)A_{\mu}(x)$ the
Abelian field strength tensor.  After a transformation to Euclidean
spacetime, we discretize the Lagrangian on a square lattice with
lattice spacing $a$ and physical extent $L \times L_t $ with periodic
boundary conditions (antiperiodic for the fermion fields in temporal
direction). We use the Wilson gauge action for the gauge field
$U_\mu \in \text{U}(1)$ and include a Wilson term in the fermion
derivative to circumvent the fermion doubling. The chemical potentials
for the two fermions are introduced by furnishing the forward and
backward temporal hopping terms with factors of $e^{\pm \mu_{u,d}}$
\cite{Hasenfratz:1983ba}, where the fermion chemical potentials
$\mu_{u,d}$ are related to the isospin chemical potential $\mu_I$ via
$\mu_u = -\mu_d = \frac{1}{2}\mu_I$.  The resulting Euclidean lattice
action decomposes into a gluonic part $S_{g}$, which contains the
dimensionless inverse coupling $\beta = 1/(ag)^2$, and two fermionic
parts (for each fermion flavour), such that
\begin{align}
 S_{E}[\bar{\psi},\psi,U,\mu_I] =
  S_{g}[U]+\bar{u}\, \text{M}[U,\mu_I]\,u +
  \bar{d} \, \text{M}[U,-\mu_I] \, d \, ,
\end{align}
where $\text{M}$ denotes the Wilson-Dirac matrix for one fermion
flavour.  Integrating out the fermionic degrees of freedom in the
grand-canonical partition function yields the determinant of the
Wilson-Dirac matrix for each flavour,
\begin{align}
 \mathcal{Z}_{GC}(\mu_I) = \int
  \mathcal{D}U\mathcal{D}\bar{\psi}\mathcal{D}\psi e^{-S_{E}} = \int
  \mathcal{D}U
  e^{-S_{g}[U]}\det\text{M}[U,\mu_I]\,\det\text{M}[U,-\mu_I] \, .
\end{align}
The fugacity expansion for a single fermion flavour separates the
isospin chemical potential $\mu_I$ from the determinants,
\begin{align}\label{eq:fugacity_exp_det}
 \det \text{M}[U,\pm \mu_I]  = \sum_{n =
  -L/a}^{L/a}\text{det}_{n} \text{M}[U] \, e^{\pm\frac{\mu_I}{T}\frac{1}{2}n},
\end{align}
where the sum over the fermion number is restricted by the lattice
volume $L/a$.  The canonical determinants $\text{det}_{n} \text{M} $
can be defined in terms of dimensionally reduced transfer matrices
with fixed fermion number $n$ and provide the projection onto the
canonical sectors with $n$ fermions
\cite{Steinhauer:2014oda}. Finally, from
\begin{align}\label{eq:Z_GC_Z_nu_nd}
    {\cal Z}_{GC}(\mu_I,T) =
  \sum_{n_u,n_d}e^{\frac{\mu_I}{T}\frac{1}{2}(n_u-n_d)}\mathcal{Z}_{n_u,n_d}(T)\, ,
\end{align}
where the dependence on the temperature $T=1/L_t$ is now made
explicit, we obtain the canonical partition functions
$\mathcal{Z}_{n_u,n_d}(T)$ given by
\begin{align}\label{eq:Z_nu_nd}
  \mathcal{Z}_{n_u,n_d}(T) = \int \mathcal{D}U
  e^{-S_{g}[U]}\text{det}_{n_u}\text{M}[U] \, \text{det}_{n_d}\text{M}[U]. 
\end{align}
The number of up and down fermions is restricted by Gauss' law. It
requires that the total electric charge $Q$, and hence the total
fermion number, has to be zero, while the total isospin charge is not restricted, i.e.,
\begin{align}
 Q = n_u + n_d = 0 \quad \quad \text{ and }\quad \quad I =
  \frac{n_u-n_d}{2} \quad \text{ arbitrary}.%\in \mathbb{Z}\, .
\end{align}
Consequently, a canonical sector with $n$ up fermions also has $n$
antidown (or equivalently $-n$ down) fermions which may bind together
to form $n$-meson states.  The collection of all states with $n$ up
and $-n$ down fermions forms the canonical partition function
$\mathcal{Z}_{n,-n}(T)$.  The vacuum sector contains flavour singlet
states and meson-antimeson states with isospin $I=0$, and is described
by the partition function $\mathcal{Z}_{0,0}(T)$.

In the canonical formalism, it is now straightforward to examine
multi-meson states and their ground-state energies $E_{n\pi}$ by
taking the free energy difference between the corresponding canonical
sector and the vacuum and extrapolating it to zero temperature
\begin{align}
 E_{n\pi}= -\lim_{T\rightarrow
  0}T\log\left(\frac{\mathcal{Z}_{+n,-n}(T)}{\mathcal{Z}_{0,0}(T)}\right).
  \label{eq:Enp definition}
\end{align}
We have explicitly checked in the $1$-, $2$- and $3$-meson sectors
that the ground-state energies coincide with the direct measurements
of the corresponding energies extracted from correlators formed with
$\pi^{+}$, $\pi^{+}\pi^{+}$ and $\pi^{+}\pi^{+}\pi^{+}$ operators.

\section{Isospin $I=1$ sector and finite-volume corrections}
\begin{figure}[!bh]
 \begin{center}
    \includegraphics[width=0.80\textwidth]{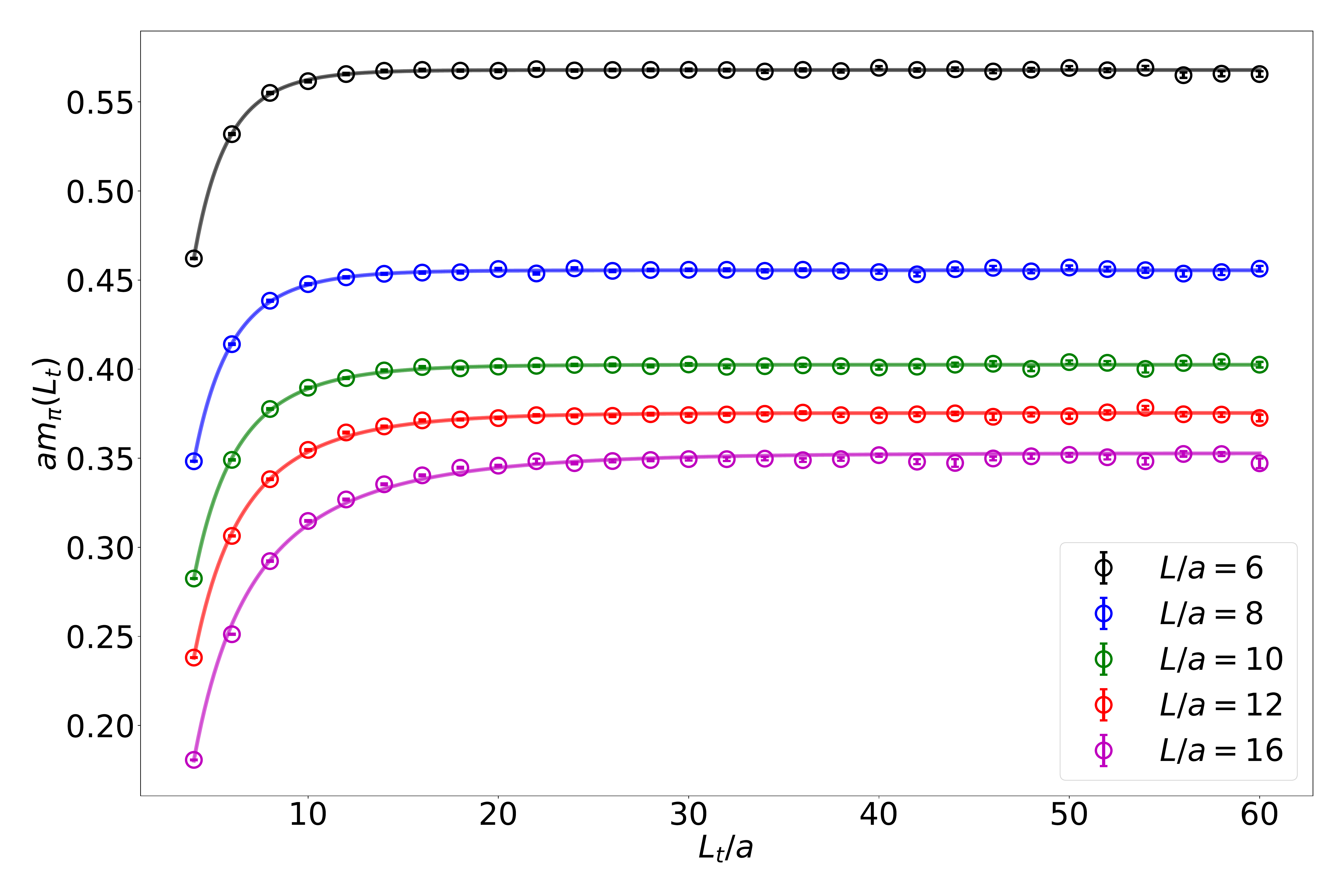}
  \vspace{-0.4cm}
    \caption{Temperature dependence of the pion mass for different
      volumes at fixed lattice spacing $\beta=5.0$. The lines with
      error bands represent fits including excited state contributions
      from the $I=1$ and the vacuum sector.}
    \label{fig:EPI_Lx_Lt}
  \end{center}
\end{figure}

The lightest particles in the massive $2$-flavour Schwinger model form
the mass-degenerate meson (or pion) triplet
$|\pi \rangle = \lbrace |\pi^{-}\rangle, |\pi^{0}\rangle,
|\pi^{+}\rangle\rbrace$.  Within this triplet, the state with maximal
$z$-component of the isospin is built from an up and antidown fermion,
$|\pi^{+}\rangle = |u \bar{d}\rangle$. It can be identified as the
ground state of the isospin $I=1$ sector described by the canonical
partition function in eq.~(\ref{eq:Z_nu_nd}) with
$\{n_u,n_d\} = \{1,-1\}$.  Hence, the ground-state energy of the
$1$-meson sector, i.e., the mass of the pion, is determined by
eq.~(\ref{eq:Enp definition}) using $n=1$.  We use this prescription
to compute the pion mass $m_{\pi}(L)$ for different volumes $L$, as
illustrated in figure \ref{fig:EPI_Lx_Lt}, where we show the behaviour
of the pion mass towards zero temperature, i.e.,
$L_t \rightarrow \infty$. It is governed by contributions from
excited states in the $I=1$ sector and the vacuum
sector. Corresponding fits to the data are shown in figure
\ref{fig:EPI_Lx_Lt} as lines with (barely visible) error bands. In the
canonical formalism excited states contribute to the free energy
differences with amplitudes given solely by their degeneracies, see,
e.g.,~Ref.~\cite{Fodor:2007ga}. This is in contrast to traditional
spectroscopy with correlation functions, where the excited state
contributions depend on the overlap of the pion operators with the
pion wave function.  The results can now be used to investigate
finite-volume effects. They arise when the wave function of the pion
overlaps at the boundaries of the box and therefore interacts with
itself. This leads to an increase of the pion mass for small volumes.
L\"uscher appropriately called these kinds of effects "interactions
around the world" and provided a formula that can be used to describe
these finite-volume effects \cite{Luscher:1985dn}. In the case of a
two-dimensional quantum field theory, one has
\begin{align}
m_\pi(L)    &=  m_{\pi}
              +\frac{1}{\sqrt{m_{\pi}L}}\left(\frac{F(0)}{\sqrt{2\pi}
              4 m_{\pi}}\right)e^{- m_{\pi}
              L}+\left(\frac{\lambda^2}{4\sqrt{3}
              m_{\pi}^3}\right)e^{-\frac{\sqrt{3}}{2} m_{\pi}L}.
              \label{eqn:FV ansatz}
\end{align}
Here, $m_{\pi}= \lim_{L\rightarrow \infty}m_{\pi}(L)$ denotes the
infinite-volume pion mass, $F(0)$ the forward scattering amplitude,
and $\lambda$ some effective 3-meson coupling.

\begin{figure}[!tbh]
%   \vspace{-0.3cm}
  \vspace{-0.39cm}
 \begin{center}
    \includegraphics[width=0.90\textwidth]{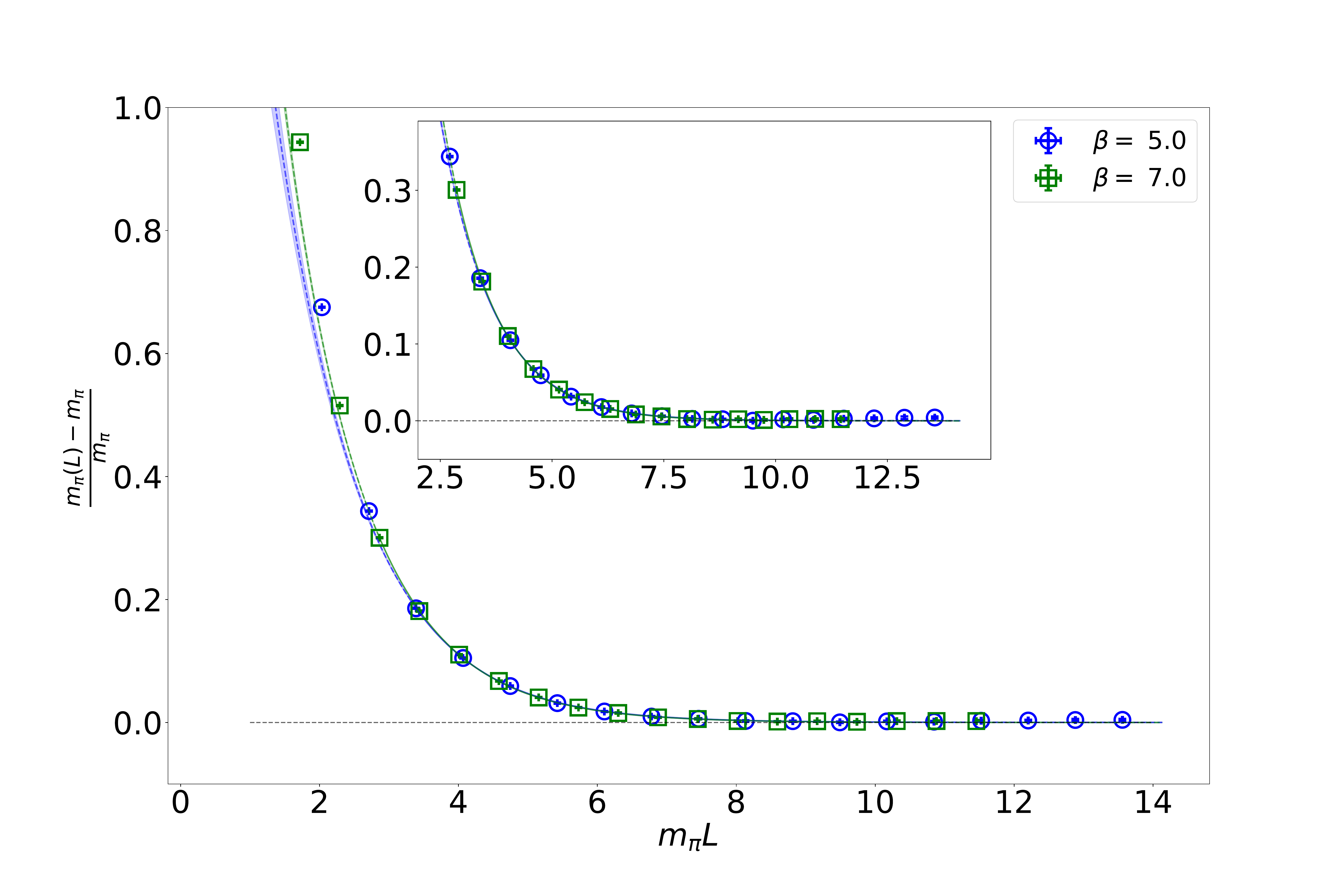}
%    \vspace{-0.3cm}
   \vspace{-0.39cm}
   \caption{Volume dependence of the pion mass $m_{\pi}(L)$ for two
      different lattice spacings $\beta=5.0$ and $7.0$ with the
      infinite-volume pion mass fixed at
      $m_{\pi}\sqrt{\beta} \sim 0.7580$. Shown are the relative
      finite-volume corrections. The lines with error bands represent
      L\"uscher's finite-volume formula in eq.~(\ref{eqn:FV
        ansatz}). }
    \label{fig:EPI_Lx}
  \end{center}
\end{figure}

In figure \ref{fig:EPI_Lx} we show the relative finite-volume effects
in the pion mass $\delta m_\pi = (m_\pi(L) - m_\pi)/m_\pi$ at two
different lattice spacings $\beta \in \lbrace 5.0,7.0\rbrace$. We use
L\"uscher's ansatz to describe the finite-volume effects using
$m_{\pi}$, $F(0)$ and the effective 3-meson coupling $\lambda$ as fit
parameters. The ansatz allows us to describe the measurements down to
small volumes $m_\pi L \gtrsim 3.0$. In order to do so, we need to
include the term related to the effective three-meson coupling. While
G-parity forbids a 3-pion coupling, an effective 3-meson coupling can
apparently not be excluded.  For the measurements presented here, we
kept the infinite-volume pion mass fixed in physical units, i.e.,
$m_\pi\sqrt{\beta} = m_{\pi}/g \sim 0.7580$, in order to estimate
lattice artefacts. Our results indicate that the artefacts are very
well under control, even for small volumes.  It is interesting to note
that the data for the relative finite-volume corrections
$\delta m_\pi$ obtained at different pion masses also fall onto the
same curve, emphasizing the universal character of the corrections
given by eq.~(\ref{eqn:FV ansatz}).

\section{Isospin $I=2$ sector and  scattering phase shifts}
\label{sec:I=2 sector}
\begin{figure}[!ht]
  \vspace{-0.4cm}
  \begin{center}
      \includegraphics[width=0.90\textwidth]{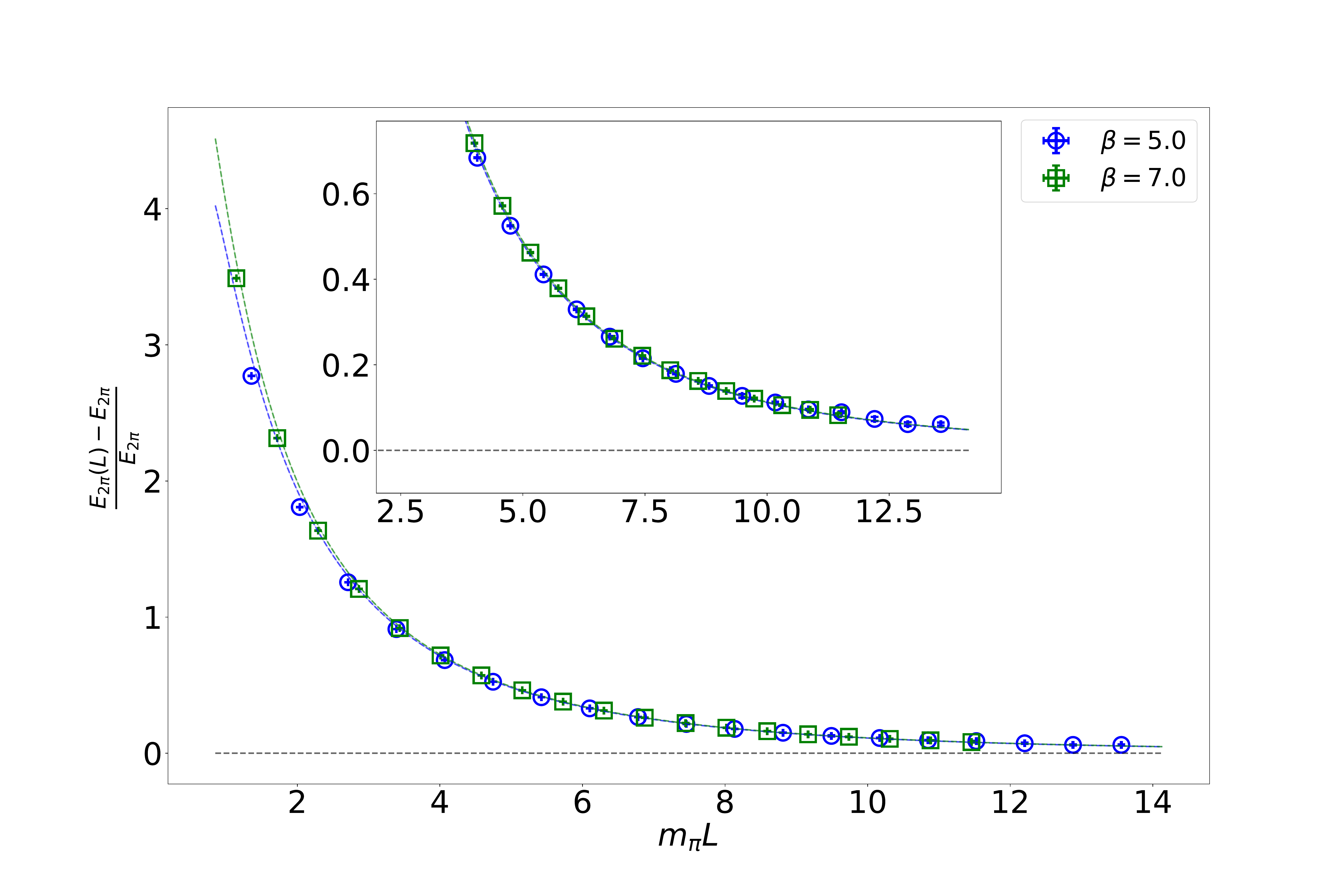}
   \vspace{-0.4cm}
    \caption{Volume dependence of the two-pion energy $E_{2\pi}(L)$
      for two different lattice spacings $\beta=5.0$ and $7.0$ with
      the infinite-volume pion mass fixed at
      $m_{\pi}\sqrt{\beta} \sim 0.7580$. Shown are the relative
      finite-volume corrections. The lines with error bands represent
      fits resulting from an effective ansatz for the scattering phase shift.}
    \label{fig:E23PI_Lx}
  \end{center}
\end{figure}

\begin{figure}[!t]
 \begin{center}
    \includegraphics[width=0.80\textwidth]{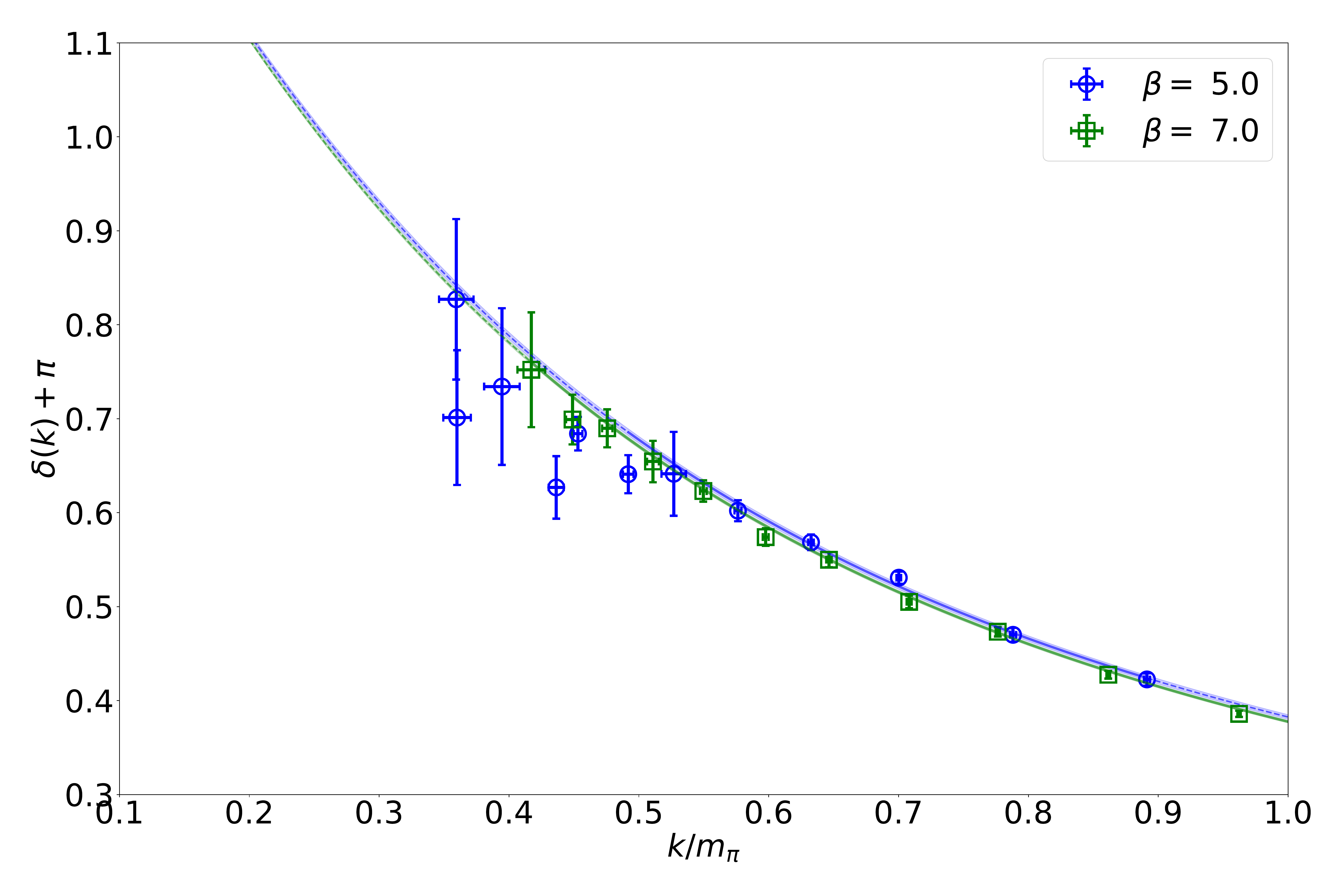}
   \vspace{-0.2cm}
    \caption{Scattering phase shift $\delta(k)$ for two different
      lattice spacings $\beta=5.0$ and $7.0$ with the infinite-volume
      pion mass fixed at $m_{\pi}\sqrt{\beta} \sim 0.7580$. The lines
      with error bands represent fits with an effective ansatz
      motivated by analytical results from the Sine-Gordon model.}
    \label{fig:Scatt}
  \end{center}
\end{figure}

The ground-state energies in the isospin $I=2$ sector, corresponding
to the energies of the 2-meson (or 2-pion) states, are obtained from
eq.~(\ref{eq:Enp definition}) with $n=2$. The results for the relative
finite-volume corrections
$\delta E_{2\pi}= (E_{2\pi}(L) - E_{2\pi})/E_{2\pi}$ are depicted in
figure \ref{fig:E23PI_Lx} as a function of the volume.
The volume dependence of the $2$-pion ground-state energies is of
particular interest in the context of scattering.  Consider a
situation where one has two pions in a box of size $L$ with equal
masses $m_{\pi}$ and momenta $p_1, p_2$.  The continuum dispersion
relation for such a state in the center of mass frame
($P = p_1+p_2 = 0$) reads
\begin{align}
  E_{2\pi} = 2 \sqrt{m_{\pi}^2+k(L)^{2}},
  \label{eq:E2PI_Boson} 
\end{align}
where $\pm k(L)$ denote the volume-dependent momenta of the two pions
in the finite box. These momenta are determined by the scattering
phase shift $\delta(L)$ which needs to be introduced due to the
boundary conditions. The phase shifts are described by the
intriguingly simple quantization condition
\begin{align}\label{eq:E2PI_Quant}
 \delta(k(L)) = -\frac{k(L) L}{2} \equiv \delta(L) 
\end{align}
as shown by L\"uscher in \cite{Luscher:1986pf}.  If the scattering
phase shift $\delta(k)$ is known, one can construct the allowed
relative momenta $k$ and compute the $2$-pion ground-state energy for
arbitrary volumes.  Conversely, one can determine the scattering phase
shifts from the 2-pion energies by using the dispersion relation
eq.~(\ref{eq:E2PI_Boson}) in combination with the quantization
condition eq.~(\ref{eq:E2PI_Quant}). In order to (partially) account
for lattice artefacts, we use the bosonic lattice dispersion relation
\begin{align}
  E_{2\pi}(L) = 2 \cosh^{-1}(\cosh(m_{\pi})+1-\cos(k(L)))
  \label{eq:E2PI_Boson_lattice}  
\end{align}
instead of the continuum one in eq.~(\ref{eq:E2PI_Quant}).  The results
are shown in figure \ref{fig:Scatt} for two lattice spacings
$\beta=5.0$ and $7.0$ and infinite-volume pion mass fixed at
$m_{\pi}\sqrt{\beta} \sim 0.7580$. The scattering phase shift can be
fitted using an effective ansatz motivated by the analytical result
from the Sine-Gordon model.\footnote{In the strong-coupling limit, the
  $2$-flavour Schwinger model goes over to the Sine-Gordon model.} In
this way, we obtain a heuristic description of the scattering phase
shift $\delta(k)$ for arbitrary $k$.
The results of the fits are
shown in figures \ref{fig:E23PI_Lx} and \ref{fig:Scatt} by the lines
with error bands.

\section{Isospin $I=3$ sector and 3-particle  quantization conditions}
\begin{figure}[!t]
   \vspace{-0.4cm}
 \begin{center}
    \includegraphics[width=0.90\textwidth]{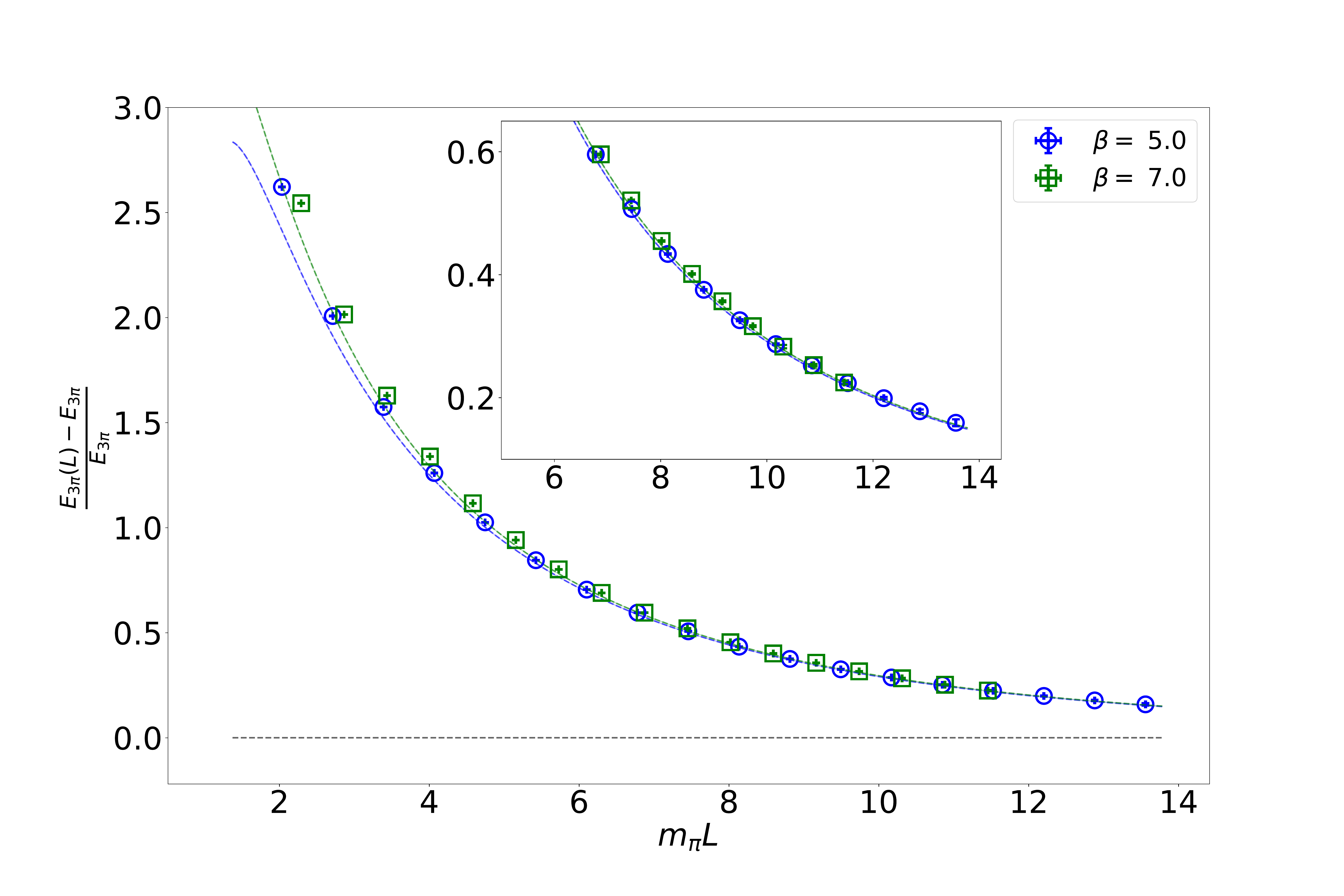}
   \vspace{-0.4cm}
   \caption{Finite-volume dependence of the 3-pion ground-state
      energies $E_{3\pi}(L)$ at two different lattice spacings
      $\beta= 5.0$ and $7.0$ with the infinite-volume pion mass fixed
      at $m_{\pi}\sqrt{\beta} \sim 0.7580$. Shown are the relative
      finite-volume corrections. The lines and error bands correspond
      to the predictions based on the 3-particle quantization
      conditions eqs.~(\ref{eq:quantization conditions Guo}) and the
      scattering phase shift obtained in Sec.~\ref{sec:I=2 sector}.}
    \label{fig:E3PI}
  \end{center}
\end{figure}

Next, we consider three pions in a finite box of size $L$ and
determine the ground-state energies in the corresponding isospin $I=3$
sector from eq.~(\ref{eq:Enp definition}) with $n=3$. In figure
\ref{fig:E3PI} we show the relative finite-volume corrections
$\delta E_{3\pi}= (E_{3\pi}(L) - E_{3\pi})/E_{3\pi}$ as a function of
the volume. The bosonic lattice dispersion relation for the energy of
a 3-pion state reads
\begin{align}\label{eq:E3PI_Boson}
    E_{3\pi}(L) =
  \sum_{i=1,2,3}\cosh^{-1}(\cosh(m_{\pi})+1-\cos(p_{i}(L))) \, ,
\end{align}
where the $p_i(L)$ denote the volume-dependent momenta of the three
pions.  Following the work in \cite{Guo:2016fgl,Guo:2018xbv} the
momenta of the three pions are determined by the 3-particle
quantization conditions based on the scattering phase shift. These
quantization conditions are valid in a nonrelativistic setup and under
the assumption that only short-ranged 2-particle interactions are
present, i.e., 3-particle interactions arise only from a sequence of
subsequent 2-particle interactions.  Of course, it is not clear to
what extent these assumptions are fulfilled in the 2-flavour Schwinger
model we consider here. In the center of mass frame, where
$P = p_1+p_2+p_3 = 0$, the 3-particle quantization conditions read
\begin{equation}
    \cot\left(\delta(-q_{31})+\delta(q_{12})\right)+\cot\left(\frac{p_1 L}{2}\right) = 0\,,\qquad
    \cot\left(\delta(-q_{23})+\delta(q_{12})\right)-\cot\left(\frac{p_2 L}{2}\right) = 0\,,
\label{eq:quantization conditions Guo}
\end{equation}
with $q_{ij} = (p_i-p_j)/2$.  These equations can now be solved using
the previously determined scattering phase shift $\delta(k)$, yielding
the momenta $p_i(L)$ and subsequently the 3-pion energy through the
bosonic dispersion relation in eq.~(\ref{eq:E3PI_Boson}).
In this way,
we obtain predictions for the 3-pion ground-state energies and the
corresponding relative finite-volume corrections based on the
quantization conditions and the scattering phase shift.
In figure \ref{fig:E3PI} we present the results of this exercise
together with our direct determinations using the free energy
differences. Shown are the relative finite-volume corrections. The
comparison demonstrates nice agreement down to surprisingly small
volumes.

\section{Summary}
In these proceedings, we reported some results concerning
finite-volume effects and meson-scattering in the $2$-flavour
Schwinger model.  Using the canonical formalism we determined the
ground-state energies in the sectors with fixed fermion numbers and
hence with fixed isospin. These energies are the energies of the
corresponding multi-pion states.  In this way, we extracted the pion
mass $m_{\pi}(L)$, as well as the 2- and 3-pion ground-state energies
$E_{2\pi}(L), E_{3\pi}(L)$ as a function of the spatial volume $L$.
The infinite-volume pion mass $m_{\pi}$ and the $2$-pion ground-state
energy were then used to compute the scattering phase shift
$\delta(k(L))$ for each volume. The momentum dependence of the phase
shift can be well described in terms of a heuristic ansatz inspired by
the Sine-Gordon model. Using the infinite-volume pion mass, the
scattering phase shift and the 3-particle quantization conditions from
\cite{Guo:2016fgl,Guo:2018xbv}, the 3-pion ground-state energies can
be predicted. The comparison between these predictions and our direct
measurements shows very good agreement down to rather small volumes.

\bibliographystyle{JHEP}
\bibliography{Bibtex}

\providecommand{\href}[2]{#2}\begingroup\raggedright\begin{thebibliography}{10}

\bibitem{Schwinger:1962tp}
J.~S. Schwinger, \emph{{Gauge Invariance and Mass. 2.}},
  \href{http://dx.doi.org/10.1103/PhysRev.128.2425}{\emph{Phys. Rev.} {\bf 128}
  (1962) 2425--2429}.

\bibitem{COLEMAN1975267}
S.~Coleman, R.~Jackiw and L.~Susskind, \emph{Charge shielding and quark
  confinement in the massive schwinger model},
  \href{http://dx.doi.org/https://doi.org/10.1016/0003-4916(75)90212-2}{\emph{Annals
  of Physics} {\bf 93} (1975) 267 -- 275}.

\bibitem{COLEMAN1976239}
S.~Coleman, \emph{More about the massive schwinger model},
  \href{http://dx.doi.org/https://doi.org/10.1016/0003-4916(76)90280-3}{\emph{Annals
  of Physics} {\bf 101} (1976) 239 -- 267}.

\bibitem{Detmold:2008fn}
W.~Detmold, M.~J. Savage, A.~Torok, S.~R. Beane, T.~C. Luu, K.~Orginos et~al.,
  \emph{{Multi-Pion States in Lattice QCD and the Charged-Pion Condensate}},
  \href{http://dx.doi.org/10.1103/PhysRevD.78.014507}{\emph{Phys. Rev. D} {\bf
  78} (2008) 014507}, [\href{http://arxiv.org/abs/0803.2728}{{\tt 0803.2728}}].

\bibitem{Guo:2016fgl}
P.~Guo, \emph{{One spatial dimensional finite volume three-body interaction for
  a short-range potential}},
  \href{http://dx.doi.org/10.1103/PhysRevD.95.054508}{\emph{Phys. Rev. D} {\bf
  95} (2017) 054508}, [\href{http://arxiv.org/abs/1607.03184}{{\tt
  1607.03184}}].

\bibitem{Guo:2018xbv}
P.~Guo and T.~Morris, \emph{{Multiple-particle interaction in (1+1)-dimensional
  lattice model}},
  \href{http://dx.doi.org/10.1103/PhysRevD.99.014501}{\emph{Phys. Rev. D} {\bf
  99} (2019) 014501}, [\href{http://arxiv.org/abs/1808.07397}{{\tt
  1808.07397}}].

\bibitem{Alexandru:2010yb}
A.~Alexandru and U.~Wenger, \emph{{QCD at non-zero density and canonical
  partition functions with Wilson fermions}},
  \href{http://dx.doi.org/10.1103/PhysRevD.83.034502}{\emph{Phys. Rev. D} {\bf
  83} (2011) 034502}, [\href{http://arxiv.org/abs/1009.2197}{{\tt 1009.2197}}].

\bibitem{Steinhauer:2014oda}
K.~Steinhauer and U.~Wenger, \emph{{Loop formulation of supersymmetric
  Yang-Mills quantum mechanics}},
  \href{http://dx.doi.org/10.1007/JHEP12(2014)044}{\emph{JHEP} {\bf 12} (2014)
  044}, [\href{http://arxiv.org/abs/1410.0235}{{\tt 1410.0235}}].

\bibitem{Hasenfratz:1991ax}
A.~Hasenfratz and D.~Toussaint, \emph{{Canonical ensembles and nonzero density
  quantum chromodynamics}},
  \href{http://dx.doi.org/10.1016/0550-3213(92)90247-9}{\emph{Nucl. Phys. B}
  {\bf 371} (1992) 539--549}.

\bibitem{Kratochvila:2005mk}
S.~Kratochvila and P.~de~Forcrand, \emph{{The Canonical approach to finite
  density QCD}}, \href{http://dx.doi.org/10.22323/1.020.0167}{\emph{PoS} {\bf
  LAT2005} (2006) 167}, [\href{http://arxiv.org/abs/hep-lat/0509143}{{\tt
  hep-lat/0509143}}].

\bibitem{Fodor:2007ga}
Z.~Fodor, K.~K. Szabo and B.~C. Toth, \emph{{Hadron spectroscopy from canonical
  partition functions}},
  \href{http://dx.doi.org/10.1088/1126-6708/2007/08/092}{\emph{JHEP} {\bf 08}
  (2007) 092}, [\href{http://arxiv.org/abs/0704.2382}{{\tt 0704.2382}}].

\bibitem{Alexandru:2005ix}
A.~Alexandru, M.~Faber, I.~Horvath and K.-F. Liu, \emph{{Lattice QCD at finite
  density via a new canonical approach}},
  \href{http://dx.doi.org/10.1103/PhysRevD.72.114513}{\emph{Phys. Rev. D} {\bf
  72} (2005) 114513}, [\href{http://arxiv.org/abs/hep-lat/0507020}{{\tt
  hep-lat/0507020}}].

\bibitem{Alexandru:2017dcw}
A.~Alexandru, G.~Bergner, D.~Schaich and U.~Wenger, \emph{{Solution of the sign
  problem in the Potts model at fixed fermion number}},
  \href{http://dx.doi.org/10.1103/PhysRevD.97.114503}{\emph{Phys. Rev. D} {\bf
  97} (2018) 114503}, [\href{http://arxiv.org/abs/1712.07585}{{\tt
  1712.07585}}].

\bibitem{Burri:2019wge}
S.~Burri and U.~Wenger, \emph{{The Hubbard model in the canonical
  formulation}}, \href{http://dx.doi.org/10.22323/1.363.0249}{\emph{PoS} {\bf
  LATTICE2019} (2019) 249}, [\href{http://arxiv.org/abs/1912.09361}{{\tt
  1912.09361}}].

\bibitem{Buhlmann:2021gzz}
P.~B\"uhlmann and U.~Wenger, \emph{{Heavy-dense QCD at fixed baryon number
  without a sign problem}},  in \emph{{38th International Symposium on Lattice
  Field Theory}}, 10, 2021.
\newblock \href{http://arxiv.org/abs/2110.15021}{{\tt 2110.15021}}.

\bibitem{Hasenfratz:1983ba}
P.~Hasenfratz and F.~Karsch, \emph{{Chemical Potential on the Lattice}},
  \href{http://dx.doi.org/10.1016/0370-2693(83)91290-X}{\emph{Phys. Lett. B}
  {\bf 125} (1983) 308--310}.

\bibitem{Luscher:1985dn}
M.~L{\"u}scher, \emph{{Volume Dependence of the Energy Spectrum in Massive
  Quantum Field Theories. 1. Stable Particle States}},
  \href{http://dx.doi.org/10.1007/BF01211589}{\emph{Commun. Math. Phys.} {\bf
  104} (1986) 177}.

\bibitem{Luscher:1986pf}
M.~L{\"u}scher, \emph{{Volume Dependence of the Energy Spectrum in Massive
  Quantum Field Theories. 2. Scattering States}},
  \href{http://dx.doi.org/10.1007/BF01211097}{\emph{Commun. Math. Phys.} {\bf
  105} (1986) 153--188}.

\end{thebibliography}\endgroup

\end{document}